\documentstyle[twocolumn,aps]{revtex}

\begin{document}
\draft
\title{Quantum Key Distribution with High Loss:
 Toward Global Secure Communication}

\author{Won-Young Hwang  \cite{email}} 

\address{Department of Electrical and Computer Engineering, 
Northwestern University, Evanston, IL 60208}

\maketitle
\begin{abstract}
We propose a decoy-state 
method to overcome the photon-number-splitting
attack for Bennett-Brassard 1984
quantum key distribution protocol in the presence of high loss:
A legitimate user intentionally and randomly replaces  signal
pulses by multi-photon pulses (decoy-states). Then they check
the loss of the decoy-states.
If the loss of the decoy-states is 
abnormally less than that of signal pulses,
the whole protocol is aborted.
Otherwise, to continue the protocol, 
they estimate loss of signal multi-photon pulses based
on that of decoy-states.
This estimation can be done with an assumption
that the two losses have similar values, that we justify.

\end{abstract}
Information processing with quantum systems enables certain tasks
that seems to be impossible with its classical counterparts, e.g.
quantum cryptography \cite{wies,bene,gold,hwan}, quantum  
computation \cite {shor}, and quantum metrologies
 \cite{dowl,delg,yue2,viol}.
In addition to the practical importance, this fact has large
theoretical and even philosophical implications.

Bennett-Brassard 1984 (BB84)
Quantum key distribution (QKD) protocol \cite{bene,gisi}
is one of the most promising quantum information processing.
It is expected that it will be the first practical quantum information
processor \cite{gisi}.

However, a bottleneck in practical realization of QKD for 
global secure communications is distance-limit: 
 Implementation of QKD has been successful at
 order of tens-kilometers \cite{gisi}. However, like in classical 
case, quantum signals are vulnerable to noise or decoherence.
 For long-distance QKD, therefore, it is desired that 
 quantum signals be amplified in the
 intermediate locations in the channel.
 However, due to the no-cloning theorem \cite{diek,woot,yue3},
 the task cannot be done in such a simple manner.
 Fortunately, however, quantum signals
can  be transported even under noisy environments to
 a remote site by quantum repeaters \cite{brie}.
However,  it is difficult to realize the
quantum repeaters with current technologies.
Therefore, we need to relax the
distance-limit in QKD without quantum repeaters.
One of the most promising candidate for this 
is to use surface-to-satellite
free-space BB84 QKD \cite{rari,hugh,kurt,klar}.
 However, the
surface-to-satellite scheme would suffer high loss. 
High loss is a serious threat to the BB84 protocol, due to
photon number splitting (PNS) attack
\cite{hutt,yuen,lutk,lut2,bras}.
Thus there have been a study \cite{lutk} and a proposal \cite{scar}
on how to overcome the PNS attack.

The purpose of this paper is to propose decoy-state
method to overcome
the PNS attack for BB84 protocol in the presence of high loss.
This paper is organized as follows.
First we will briefly review how PNS attack renders the BB 84 
protocol with high loss insecure. 
Next we decribe the decoy-state method. 
Next we argue why we can assume that the loss of decoy-states is
similar to that of other signal pulses. 
Then we derive a condition for security of the proposed
protocol. Finally we discuss and conclude.

Let us briefly describe the PNS attack \cite{hutt,yuen,lutk,lut2,bras}.
Unless perfect single-photon sources are used, 
BB84 protocol with loss is 
vulnerable to the following attack of Eve (an eavesdropper).
 Let us assume that 
 Alice (one legitimate participant) uses the following photon
sources in BB84 protocol \cite{bene}. 
Emits a pulse that contains a single-photon with, for example,
90 \% probability, and emits a pulse that contains
multi-photons with 10 \% probability.
The problem here is that multi-photons are inadvertantly generated
and thus we do not know when they have been emitted.
Also assume that the
channel loss $l$ is, for example, 90 \%, or it has 10 \% yield $y$.
Here $y=1-l$ and $y$ corresponds to $p_{exp}$ in Ref. \cite{bras}.
Here we assume that Bob (the other legitimate participant)
 uses more practical detectors 
that are insensitive to photon numbers.
A more rigorous definition of the yield will be given later.
 Eve's attacking method is the followings. 
 First, Eve measures the number of photons of each pulse.
 When it is one, she just blocks it. When it is more than
 one, she splits the photons.
(Our discussion here is valid for general photon
sources because "Alice can dephase the states to create 
a mixture of Fock states \cite{lut2,bras}", as described later.)
 Then she preserves one and sends
 the other photons via an ideal lossless channel to Bob. As usual,
 we assume that Eve has unlimited technological and computational
 power. She is only limited by laws of Nature.
Then what Bob observes is that only 10 \% of photon pulses 
arrive at him, as expected. However,
Eve can get full information about the key by measuring each of
the preserved photons in a proper basis that is publically announced
later by Alice.
We adopt the best case assumption for Eve as usual,
that all multi-photon pulses were used for the PNS attacks.
Then we can see that if the yield $y$ is less than
the probability $p_{multi}$ of multi-photon generations 
then the scheme
is totally insecure due to the PNS attacks. 
In other words,  yield $y$ must be greater than
 the probability $p_{multi}$ of multi-photon generations in order that 
the scheme be secure \cite{bras},
\begin{equation}
\label{a}
 y > p_{multi}.
\end{equation}
Here the probability $p_{multi}$ of multi-photon 
generations is a parameter for quality of (imperfect) single-photon
sources. The smaller it is, the higher the quality is.
For a  single-photon source with a given quality,
the loss that can be tolerated is determined by 
Eq. (\ref{a}).

Therefore, when yield $y$ is very low, almost
perfect single photon generator
is required \cite{lut2,bras}.
This fact motivated current developements of single-photon
sources and demonstration of QKD with them \cite{beve}. 
However, the problem is that the sources
cannot be perfect single-photons in practice \cite{waks}.
In the case of surface-to-surface free-space BB84 protocol,
this condition seems to be barely satisfied \cite{hugh}.
However, the loss in surface-to-satellite scheme 
must be higher than that in the surface-to-surface scheme,
because in the former case one party is moving fast and
more far apart. Moreover, security of surface-to-satellite
protocol in which a satellite plays a role of
 a legimate user is based on physical
security of the satellite. That is, we must assume that Eve 
cannot secretly observe inside of the satellite. However,
it is not easy to justify the assumption.
Thus we propose to use (possibly geostationary) 
satellite with mirrors that connect users on the surface.
 (The refection-protocol has been proposed
in the context of multi-users QKD \cite{klar}.) 
 In this case, it is clear that the loss will be much higher.
This means that it is difficult to implement secure
 scheme with current technologies.

Thus we need methods to directly detect the PNS attack.
One possibility is to monitor photon-number statistics that might
have been disturbed by Eve's PNS attack.
This possibility has been studied in the case of weak coherent states
\cite{lutk}; Simple minded PNS
attack disturbs the 
photon-number statistics of the pulses thus
it can be detected by Bob.
However, Eve can launch a sophisticated PNS
attack that preserves the photon-number statistics \cite{lutk}.  
Thus we need another method against the PNS attack.

The basic idea of the decoy-state method is the followings.
In PNS attack, Eve selectively transports subsets of 
multi-photons to Bob.
Thus the yield of multi-photon pulses must be abnormally higher
than that of single photon pulses.
Let us assume that Alice had intentionally and randomly 
replaces photon pulses from signal sources 
by multi-photon pulses (the decoy states).
 Since Eve cannot distinguish
multi-photon pulses of signal source from those of decoy
source, the yields of the two pulses must be similar.
Thus Alice and Bob can detect the PNS attack by checking the yield
of decoy source.

Before we give the decoy-state method, let us describe
preliminaries more precisely. Let us consider
a source that emitts a pulse $|n\rangle$ that contains
$n$ photons (in the same polarization state) 
with a probabity $p_n$.
Here $n= 0,1,2,...$  and $\sum_n p_n=1$.
Each pulse is used to encode one bit of key.
Let us also consider a source that  
generates a coherent state $|\mu e^{i\theta} \rangle$
\cite{lut2,bras,scar}.
By randomizing the phase $\theta$, the state reduces to
a mixed state $\rho=\int (d\theta/2\pi) |\mu e^{i\theta} \rangle
\langle \mu e^{i\theta}|$. However, this state is equivalent to 
mixture of Fock state $\sum_n P_n(\mu) |n\rangle \langle n|$, with
Poissonian distribution $P_n(\mu)= e^{-\mu} \mu^n/n!$.
In other words, the source that emits pulses in
coherent states $|\mu e^{i\theta} \rangle$ is, after 
phase randomization, equivalent to
a source that emits an n-photon state
$|n\rangle$ with a probability $P_n(\mu)$.

Alice adopts two photon
sources, that is, singal source $S$ and decoy source $S^{\prime}$.
Signal source is used to distribute key. Decoy source is used
to detect the PNS attack.
Let us first
 consider the most practical case where both sources
are generators of coherent states.
For signal source $S$, we adopt $\mu<1$, that is, it mostly emitts
single-photon pulses. 
For decoy source $S^{\prime}$, we adopt $\mu^{\prime} \geq 1$, 
that is, it mostly emitts multi-photon pulses. 
The polarization of the pulses of the decoy source is randomized 
such that it cannot be distingushied from those of the signal source
as long as photon numbers of the pulses are the same.

We assume that Bob uses practical photon detectors
that are insensitive to photon numbers.
The yield $y_n$ and $y_n^{\prime}$ are relative frequencies
 that  n-photon pulses from the signal and decoy sources
 are registered by Bob's detector, respectively. 
Here  $0 \leq y_n, y_n^{\prime} \leq 1$. 
It is notable that the yields can be unity even if some photons in 
a pulse are lost. It is because Bob's detector does not count 
number of 'lost photons' in a pulse that is detected.
Yield of signal source $Y_s$ and that of the decoy source
$Y_d$ are, respectively, given by
\begin{equation}
\label{b}
Y_s=\sum_n P_n(\mu) y_n, \hspace{1cm}
Y_d=\sum_n P_n(\mu^{\prime}) y_n^{\prime}.
\end{equation} 
Here $Y_s$ and $Y_d$ can be directly detected by Bob.
We also consider yield of only multi-photon pulses from singal source,
$Y_s^m$, that is given by,
\begin{equation}
\label{c}
Y_s^m=\sum_{n=2}^{\infty} P_n(\mu) y_n.
\end{equation}
This quantity cannot be directly measured but it can be bounded 
based on other yields.
Normalized yield of multi-photon pulses from singal source,
$\tilde{Y}_s^m$, is given by,
\begin{equation}
\label{d}
\tilde{Y}_s^m= \sum_{n=2}^{\infty} P_n(\mu) y_n
                 /\sum_{n=2}^{\infty} P_n(\mu).
\end{equation}
Now let us describe the protocol more precisely.
 In the decoy-state method, Alice performs BB84 protocol with
signal source $S$. However, Alice randomly replaces the
signal source $S$ by the decoy source $S^{\prime}$ with a probability
$\alpha$. 
After Bob announces that he has received all photon pulses,
Alice announces which pulses are from the decoy source.
By public discussion, they estimate the total yield of
signal source $Y_s$ and that of decoy source $Y_d$.
If $Y_d$ is too larger than $Y_s$, 
they abort the whole protocol.
Otherwise, they continue the protocol by estimating
yield of multi-photon pulses from singal source, $Y_s^m$, 
using the yield of decoy source $Y_d$ in the following way.

Eve cannot distinguish multi-photon pulses of singal source from
those of decoy source.
Thus we can expect that
the normalized yield of multi-photon pulses from signal source
is similar to that of decoy source that mainly composes
of multi-photon pulses.
Let us discuss it in more detail.
First, we can see that the yields $y_n$ and $y_n^{\prime}$ cannot
be different. That is,
\begin{equation}
\label{e}
 y_n= y_n^{\prime}.
\end{equation}
 It is because, for a given pulse with a certain photon
number, Eve can get no more information about which source
the pulse is from, than what she obtains
from the Bayes's law.
The only way for Eve to take advantage of the Bayes's law is to
control values of $y_n$ as she like.

From Eqs. (\ref{b}) and (\ref{e}), it is easy to obtain
\begin{equation}
\label{f}
\sum_{n=2}^{\infty} P_n(\mu^{\prime}) y_n
\leq Y_d.
\end{equation}
Now the problem is how 
$Y_s^m=\sum_{n=2}^{\infty} P_n(\mu) y_n$ is bounded by 
Eq. (\ref{f}). 
Eve's goal is to make
$Y_s^m$ as large as possible, for a given
yield of decoy source $Y_d$. 
In other words, it is to make 
the ratio $A \equiv \sum_{n=2}^{\infty} P_n(\mu) y_n/
\sum_{n=2}^{\infty} P_n(\mu^{\prime}) y_n$ as large as possible.
Let us now note that, for $\mu<\mu^{\prime}$ as we will choose,
\begin{equation}
\label{fb}
\frac{P_n(\mu)}{P_n(\mu^{\prime})} > 
\frac{P_m(\mu)}{P_m(\mu^{\prime})},
\hspace{5mm}
\mbox {if}
\hspace{5mm} n <m .
\end{equation}
It is because
$P_n(\mu)/P_n(\mu^{\prime})
 =  [e^{-\mu}\mu^n/n!]  /  [e^{-\mu^{\prime}} {(\mu^{\prime})}^n/n!]
 =  [e^{-\mu}  / e^{-\mu^{\prime}}] [{(\mu/\mu^{\prime})}^n] $.
We can see that
the ratio $A$  is bounded as,
\begin{equation}
\label{fc}
A = \frac{\sum_{n=2}^{\infty} P_n(\mu) y_n}
      {\sum_{n=2}^{\infty} P_n(\mu^{\prime}) y_n}
      \leq \frac{P_2(\mu)}{P_2(\mu^{\prime})}.
\end{equation}
It is because
$ P_2(\mu)/P_2(\mu^{\prime})-A
=[1/\{P_2(\mu^{\prime})
 \sum_{n=2}^{\infty} P_n(\mu^{\prime}) y_n\}]
   [P_2(\mu) \sum_{n=2}^{\infty} P_n(\mu^{\prime}) y_n
  -P_2(\mu^{\prime})\sum_{n=2}^{\infty} P_n(\mu) y_n ] 
=[1/\{P_2(\mu^{\prime}) \sum_{n=2}^{\infty} P_n(\mu^{\prime}) y_n\}]
 \sum_{n=2}^{\infty}  y_n 
 \{P_2(\mu) P_n(\mu^{\prime})-P_2 (\mu^{\prime}) P_n(\mu)\} 
 \geq 0.$
 Here we have used  Eq. (\ref{fb}) and $y_n, P_n \geq 0$.
The equality is achieved when $y_2 > 0$ and
$y_i=0$, where $i=3,4,5, ...$ .

Thus this is Eve's best choice.
This can be interpreted as follows. 
The larger the number of photons of a given pulse is,
the more probable it is that the pulse is from the decoy source,
by the Bayes's law and the property of the Poissonian distribution.
Eve had better not to make Bob's detector register when it is more
probable that the pulse is from the decoy source.
Thus Eve's optimal choice is to block pulses containing more than 2 
photons.  

By Eqs. (\ref{c}), (\ref{f}) and (\ref{fc}), we can get
\begin{equation}
\label{g}
Y_s^m \leq  \frac{P_2(\mu)}{P_2(\mu^{\prime})} Y_d.
\end{equation}
The normalized one, $\tilde{Y}_s^m$, is given by,
\begin{equation}
\label{h}
\tilde{Y}_s^m \leq \frac{1}{P_2(\mu^{\prime})}
\frac{P_2(\mu)}
      {\sum_{n=2}^{\infty} P_n(\mu)} Y_d .
\end{equation}
 $P_2(\mu^{\prime})$ and
 $P_2(\mu)/ \sum_{n=2}^{\infty} P_n(\mu)$ are
of orders of unity in reasonable regions of
$\mu$ and $\mu^{\prime}$, for example,
$\mu= 0.5$ and $\mu^{\prime}=1$.
Thus we get the expected
 result that $\tilde{Y}_s^m$ and $ Y_d$ have the same order
of magnitudes.

The condition for security in Eq. (\ref{a}) expresses
the following. 
In order that the protocol be secure, the total number of 
pulses that are detected must be greater than that of attacked ones.
In the case of the decoy-state method, the number of attacked
pulses is $[\sum_{n=2}^{\infty}P_n(\mu)] \tilde{Y}_s^m $.
Thus the condition reduces to
\begin{equation}
\label{i}
Y_s > \mbox{max} \{ [\sum_{n=2}^{\infty}P_n(\mu)] \tilde{Y}_s^m \},
\end{equation}
where the maximun is taken all strategies by Eve.
From Eqs. (\ref{h}) and (\ref{i}), we obtain a condition for security
\begin{equation}
\label{j}
Y_s > \frac{P_2(\mu)}{P_2(\mu^{\prime})} Y_d.
\end{equation}
Let us roughly estimate the quantities in our case where
both signal and decoy sources are generators of coherent states
with Poissonian statistics $P_n(\mu)$ and $P_n(\mu^{\prime})$ 
of photon numbers $n$, respectively.
In normal operations of the protocol, that is,
when Eve does not disturb the communication,
$Y_d$ will be larger than $Y_s$ by a factor of $\mu^{\prime}/\mu$.
Then the condition reduces to
\begin{equation}
\label{k}
\frac{P_2(\mu)}{P_2(\mu^{\prime})} \frac{\mu^{\prime}}{\mu}
= \frac{e^{\mu^{\prime}}}{\mu^{\prime}}\frac{\mu}{e^\mu}
<1.
\end{equation}
For a given $\mu^{\prime}$, Eq. (\ref{k}) is satisfied 
when $\mu$ is small enough, because $\mu/e^\mu < \mu $.
For example, when $\mu=0.3$ and $\mu^{\prime}=1.0$ 
the lefthand-side of Eq. (\ref{k}) is around 0.604.

One might say that the mean photon number of signal source, 
$\mu$, should still be quite smaller than unity even with decoy-state
method and thus there is no improvement over usual 
protocols without decoy-state method.
 However, this is not the case:
Eq. (\ref{k}) does not 
contain a term that amounts to channel loss,
 in contrast with the case of
Eq. (\ref{a}). Thus the condition for security 
can be satisfied no matter how high the loss is
in the normal opertations.  

Our analysis above
can be generalized to sources with any probability distribution
$p_n$. 
For example, let us consider 
almost perfect single photon generator with a particular photon
number distribution $p_1=1-\epsilon$ and $p_i =k/i! $
where $\epsilon \ll 1$, $k$ is a certain constant
satisfying $\sum_{i=2}^{\infty} p_i= \epsilon$,
and $i=2,3,...$ .
Using the same decoy source $S^{\prime}$, the condition for
security in this case is
\begin{equation}
\label{l}
Y_s > \frac{\epsilon}{P_2(\mu^{\prime})} Y_d.
\end{equation}
Eq. (\ref{l}) is satisfied by a large margin when $\epsilon$ is small
enough. However, key generation rate would be proportional to the 
margin. Thus, key generation rate can be larger than that of the case
where weak-coherent-states are used as signal source.

In general, the more similar forms the photon-number statistics of
the signal source and decoy source
have in a region of multi-photons, the more efficient the decoy-
state method is. 
Let us also consider the following extreme case.
Consider an almost-perfect
 single-photon generator with a photon number
distribution, $p_1=1-\epsilon$ and $p_{N}=\epsilon$, where
$\epsilon \ll 1$ and $N$ is a number quite larger than 2, for 
example, 10. In this case, Eve selectively attacks the $N$ photon
pulses, making it more difficult to satisfy the corresponding
security condition, if we adopt the same decoy source 
$S^{\prime}$ with Poissonian distribution $P(\mu^{\prime})$.
Thus a general stategy in a design of a pair of signal and 
decoy sources is to make the forms of the photon number
statistics of the two sources as similar as possible, in the region
of multi-photons.

The proposed method is based on the basic idea of random sampling.
Thus we believe that the security of the proposed protocol 
against the most general attacks can be shown later, possibly
extending methods developed in recent literatures
\cite{inam,koas,gott}.

Conditions for security of protocols with more practical settings,
e.g.  non-zero dark count rate and misalignment of basis, should
also be analyzed later.

The only way to address certain imperfection so far is to assume
the best-case for Eve \cite{inam,koas,gott}. For non-zero 
error rates, for example, we assume that it is entirely due to Eve's
attack.
Decoy-state method, however, is an example where we can
relax this kind of assumptions without loss security in a proper way.
It will be worthwhile to look for similar ideas that can 
address other imperfections.

In conclusion, we have proposed a decoy-state 
method to overcome the photon-number-splitting
attack for BB84
QKD protocol in the presence of high loss:
A legitimate user intentionally and randomly replaces the signal
pulses by multi-photon pulses (the decoy states).
Then they check the yield of the decoy-states.
If the yield of the decoy-states is 
abnormally higher than that of other signal pulses,
the whole protocol is aborted.
Otherwise, to continue the protocol, 
they estimate yield of signal multi-photon pulses based
on that of decoy-states. 
This estimation can be done with an assumption
that the two losses have similar values, that we justified.
We have demonstrated that the estimation can be made indeed
in the pratical case of coherent pulses sources.
However, the analysis can be generalized to arbitrary case.

I am very grateful to Prof. Horace Yuen for 
encouragement. I am also grateful to Prof.
Norbert L\"utkenhaus for helpful comments.

\end{document}